\documentclass[conference]{IEEEtran}
\usepackage{blindtext}
\usepackage{graphicx}
\usepackage{color}
\usepackage{gensymb}

\ifCLASSINFOpdf
\else
\fi
\begin{document}
%
\title{A Microservice-enabled Architecture for Smart Surveillance using Blockchain Technology}

\author{
\IEEEauthorblockN{Deeraj Nagothu, Ronghua Xu, Seyed Yahya Nikouei, Yu Chen}
\IEEEauthorblockA{Dept. of Electrical \& Computer Engineering,
Binghamton University, SUNY,  Binghamton, NY 13902, USA \\
\{dnagoth1, rxu22, snikoue1, ychen\}@binghamton.edu}
}

\maketitle

\begin{abstract}
While the smart surveillance system enhanced by the Internet of Things (IoT) technology becomes an essential part of Smart Cities, it also brings new concerns in security of the data. Compared to the traditional surveillance systems that is built following a monolithic architecture to carry out lower level operations, such as monitoring and recording, the modern surveillance systems are expected to support more scalable and decentralized solutions for advanced video stream analysis at the large volumes of distributed edge devices. In addition, the centralized architecture of the conventional surveillance systems is vulnerable to single point of failure and privacy breach owning to the lack of protection to the surveillance feed. This position paper introduces a novel secure smart surveillance system based on microservices architecture and blockchain technology. Encapsulating the video analysis algorithms as various independent microservices not only isolates the video feed from different sectors, but also improve the system availability and robustness by decentralizing the operations. The blockchain technology securely synchronizes the video analysis databases among microservices across surveillance domains, and provides tamper proof of data in the trustless network environment. Smart contract enabled access authorization strategy prevents any unauthorized user from accessing the microservices and offers a scalable, decentralized and fine-grained access control solution for smart surveillance systems. 

\end{abstract}


%
\IEEEpeerreviewmaketitle

\section{Introduction}
\label{sec:intro}
The proliferation of the Internet of Things (IoT) technology allows the concept of Smart Cities become feasible and smart surveillance is one of the most actively studied topics in the community. However, the IoT-based smart surveillance faces many challenges. On the one hand, relying on a centralized cloud computing center inevitably incurs uncertain latencies and poses extra workload to the communication networks. On the other hand, while a fog/edge computing based system is able to meet the requirements raised by delay-sensitive, mission-critical applications \cite{chen2017enabling}, \cite{nikouei2018lcnn}, \cite{xu2018real}, new challenges are also introduced by the distributed, cross-domain features, such as scalability, heterogeneity and interoperability.


In order to tackle the challenges in building large-scale, distributed smart applications and platforms for smart cities, the microservices architecture has emerged and gained a lot of popularity in recent years \cite{krylovskiy2015designing}. Compared to the traditional service oriented architectures (SOAs) that deploys the system as a monolithic unit, the microservices architecture divides an application into multiple atomic services. Each service only carries out one specific task and requires lightweight communication with other services or components of the system. It can be deployed on either single machine or multiple distributed hosts. Thus, microservices is ideal to build a service platform for cross-domain applications, like smart surveillance. 

The microservices architecture possesses many attractive features, such as scalability, fine granularity, loose coupling, continuous delivery, etc. While it is a natural match to the inherent rationale of edge-fog computing paradigm, microservices architecture demonstrates vulnerabilities in security due to its usage of distributed data and interfaces \cite{yu2018survey}. A smart surveillance system based on a microservices architecture can be deployed in a distributed network environment including an extraordinary large number of IoT devices with high heterogeneity and dynamics. A more scalable, flexible and decentralized security mechanism to address the vulnerabilities in data and communication is required. As the fundamental protocol of Bitcoin \cite{nakamoto2008bitcoin}, the blockchain protocol has been recognized as the potential to revolutionize the fundamentals of IT technology due to its many attractive characteristics such as supporting decentralization and anonymity maintenance \cite{ouaddah2016fairaccess}. 

In this position paper, a microservice architecture is proposed to enable a scalable, maintainable and secure smart surveillance system. Surveillance functions such as object detection, tracking and features extraction, are implemented as cooperative microservices, which are deployed independently and executed within individual processes. It provides a flexible, testable and maintainable framework in development and deployment. The blockchain and smart contract provide a decentralized security mechanism to protect data and to synchronize data of the communication channel. It also allows to enforce permissions on the data access by deploying smart contracts.  

The rest of this paper is organized as follows. Section \ref{sec:back_know} provides the background knowledge. The architecture and rationale of prototype design are presented in Section \ref{sec:bcms}, and Section \ref{sec:conclusion} concludes this paper with our ongoing work.


\section{Background Knowledge}
\label{sec:back_know}

\subsection{Microservices Architecture}
Traditional service-oriented architecture (SOA) is monolithic though constituting different software features in a single interconnected and interdependent application and database. While the tightly coupled dependence among functions and components enables a single package, such a monolithic architecture lacks the flexibility to support continuous development and continuous delivery, which is critical in today's quickly change, highly heterogeneous environment.
Each microservice is an individual process dedicated to certain function of the application. The microservices architecture is a decentralized architecture which constitutes of multiple such microservices. 
The individual microservices communicate with each other through the HTTP REST or a message bus asynchronously. The flexibility of microservices enables continuous, efficient, and independent deployment of application function units. Significant features of microservices include fine granularity, which means each of the microservices can be developed in different frameworks like programming languages or resources, and loose coupling where the components are independent of each other’s deployment and development \cite{yu2018survey}.

\begin{figure}[t]
    \centering
        \includegraphics[width=0.425\textwidth]{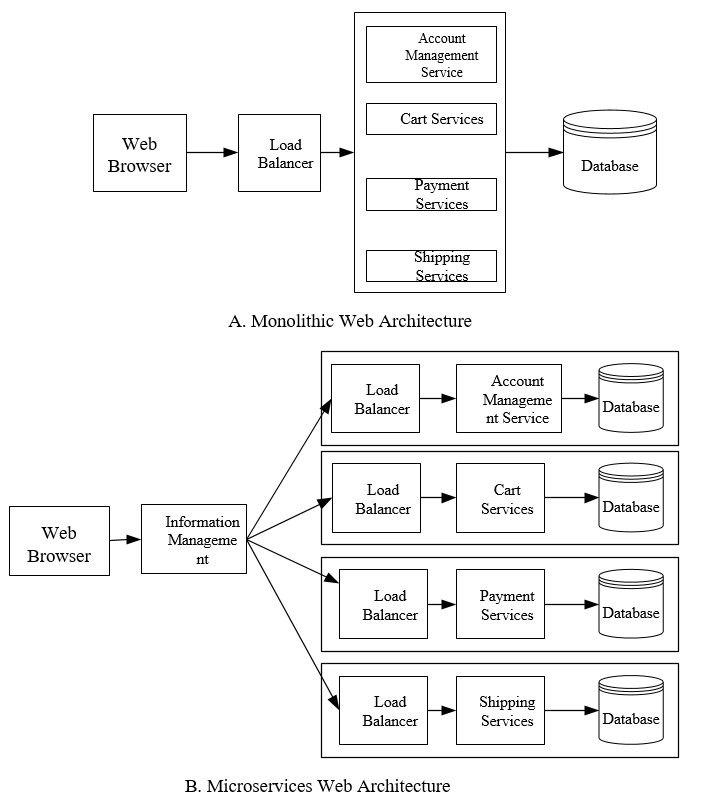}
    \caption{Commerce web based on Monolith and Microservices Framework.}
    \label{fig:micro}
    \vspace{-10pt}
\end{figure}

Microservices architecture has been adopted to revitalize the monolithic architecture based applications, including the modern commercial web application. Figure \ref{fig:micro} shows the differences between the monolithic design and the microservices architecture. The monolithic version of web design includes services like managing an individual account, cart services, payment services and shipping services. All the information is accessible and stored in a single database. If any of the services stop functioning, the rest of the information flow is halted. The service update in monolithic system requires restarting the whole application by creating a new instance of every service layer. In contrast, a web application based on the microservices architecture allows all the services run as an individual instance where the data is stored in a dedicated database for each service. This eliminates the dependability among the services. Once a service fails, it can be updated and restarted without interrupting the functionality of the entire application. 

Microservices architecture has been investigated in more smart solutions to enhance the scalability and security of applications. It was used to implement an intelligent transportation system that incorporates and combines IoT to help transportation planning for bus rapid system \cite{herrera2018smart}. The microservices architecture was used to develop a smart city IoT platform where each microservice is regarded as an engineering department. The independent behavior of each microservice allows flexibility of selecting the development platform, and the communication protocols are simplified without requiring a middleware \cite{krylovskiy2015designing}. 

\subsection{Blockchain and Smart Contract in IoT}

The blockchain, which was introduced by Nakamoto in 2008 \cite{nakamoto2008bitcoin}, has demonstrated its success in decentralization of digital currency and payment, like bitcoin. It is a public ledger that provides a verifiable, append-only chained data structure of transactions. By allowing the data be stored and updated distributively, the blockchain is essentially a decentralized architecture that does not rely on a centralized authority. The transactions are approved by a large amount of distributed nodes called miners and recorded in timestamped blocks, where each block is identified by a cryptographic hash and chained to preceding blocks in a chronological order. Blockchain uses consensus mechanism, which is enforced on miners, to maintain the sanctity of the data recorded on the blocks. Thanks to the “trustless” proof mechanism running on miners across networks, users can trust the system of the public ledger stored worldwide on many different nodes maintained by ''miner-accountants'', as opposed to having to establish and maintain trust with a transaction counter-party or a third-party intermediary \cite{swan2015blockchain}. Thus, Blockchain is an ideal decentralized architecture to ensure distributed  transactions between all participants in a trustless environment, like IoT networks.

Because of these attractive characteristics, researchers have verified the feasibility of using blockchain technique to address security issues in the IoT networks. Emerging from the smart property, smart contract allows users to achieve agreements among parties and supports variety of flexible transaction types through blockchain network. By using cryptographic and security mechanisms, smart contract combines protocols with user interfaces to formalize and secure relationships over computer networks~\cite{szabo1997formalizing}. A smart contract includes a collection of pre-defined instructions and data that have been saved at a specific address of blockchain as a Merkle hash tree, which is a constructed bottom-to-up binary tree data structure. Through exposing public functions or application binary interfaces (ABIs), a smart contract interacts with users to offer predefined business logic or contract agreement. The smart contract enabled security mechanism for IoT systems has been a hot topic and some efforts have been reported recently, for example, data protection \cite{lee2018implementation} and access control \cite{xu2018blendcac, zhang2018smart, xu2018smartcac}. We believe blockchain and smart contract together are promising to provide a solution to enable secured data sharing and access control in distributed microservices enabled smart surveillance systems.

\section{Blockchain Integrated Microservices Architecture}
\label{sec:bcms}

\begin{figure}[t]
    \centering
        \includegraphics[width=0.425\textwidth]{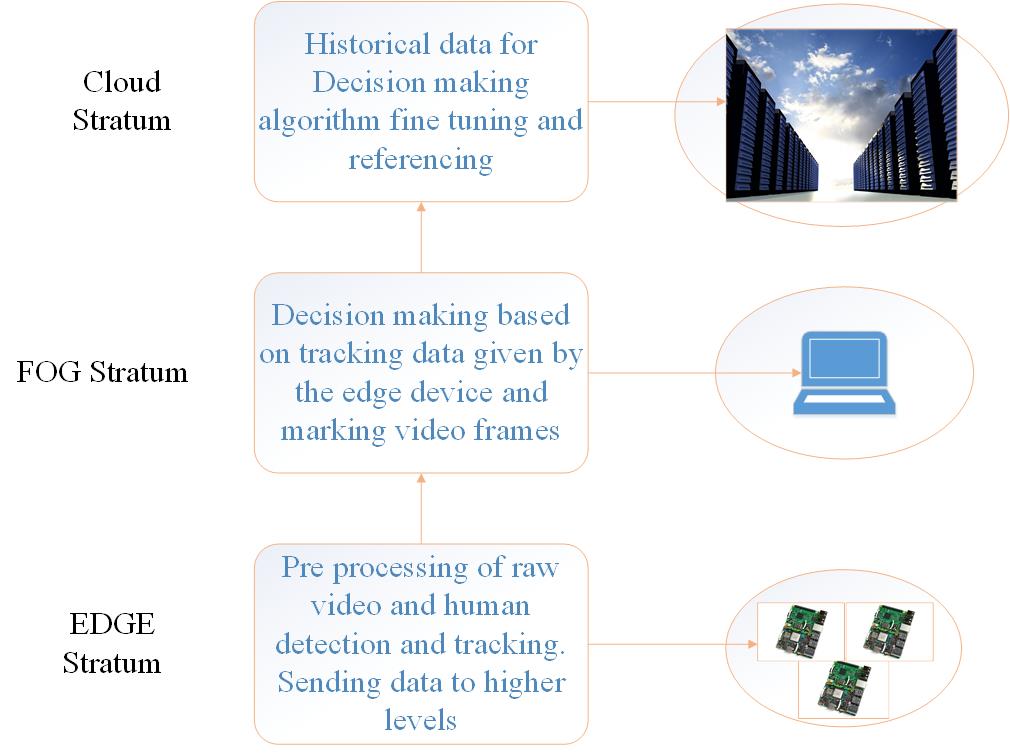}
    \caption{Layered smart surveillance system hierarchy using the edge-fog-cloud computing paradigm.}
    \label{fig:implementation}
    \vspace{-10pt}
\end{figure}

In this position paper, a novel blockchain-enhanced microservices architecture is introduced for the smart surveillance system, which is based on a hierarchical edge-fog-cloud computing paradigm as shown by Fig. \ref{fig:implementation} \cite{xu2018real}.

\subsection{Microservices Synchronization on Blockchain Network}

Edge-based video analysis framework becomes popular in the current IoT technology. The video analysis process includes various functions like real-time behavioral analysis of subjects, license plate recognition, face recognition and gesture analysis. However, running these complex algorithms at a single node requires high computational resources and storage space. In addition, updating an algorithm requires recompiling the whole source code, reconfiguring application and rebooting the entire system. Furthermore, the analytical results generated from different algorithms are stored in a single centralized database which is vulnerable as a single point of failure. It may lead to losing video footage or raising the false positive alarm rate.

\begin{figure}[t]
    \centering
        \includegraphics[width=0.425\textwidth]{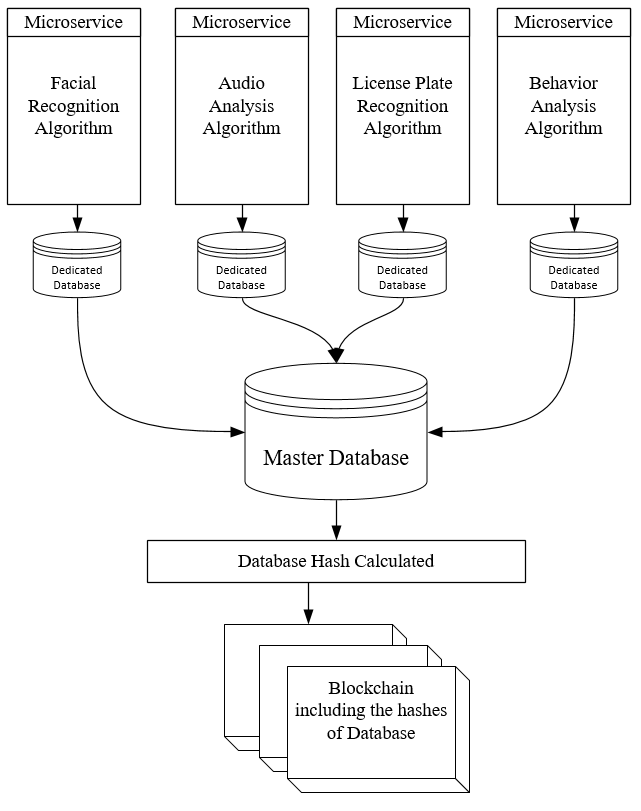}
    \caption{Block diagram of video analysis microservices on blockchain.}
    \label{fig:micro_idea1}
    \vspace{-10pt}
\end{figure}

Decoupling the complicated video analysis functions into distributed microservices reduces the inter-dependability among different algorithm execution. As shown by Fig. \ref{fig:micro_idea1}, for instance, the key algorithms in video analysis application can be developed separately as four microservices, which are deployed either on a single fog node or multiple distributed edge/fog nodes. Each microservice is dedicated to the defined function of the system and uses a private database. The microservices architecture allows each algorithm to run and update independently without interrupting the functionalities of other parts of the application.

The data saved in multiple microservice owned databases are collected at the fog level database called master database. These fog level databases use the blockchain technology to synchronize data in the trustless network environment. The event-based analysis results saved on master databases are calculated as hash value and recorded on blockchain network after proof of work done by the miners. The consensus mechanism enforced by blockchain audits the data recored on the blocks to prevent data tampering. Due to the security characteristics provided by blockchain, extracted information from each microservice will be securely synchronized among distributed fog level databases without relying on third party authority to maintain the trust relationship.

\subsection{Smart Contract Enabled Microservices System}

In a smart surveillance system shown in Fig. \ref{fig:implementation} that is implemented at the edge of the network, each surveillance camera is connected to or function as an edge device. The real time video stream will be processed on-site or near-site at the edge or fog devices instead of being sent to the remote cloud. Each object of interest is detected and marked using lightweight algorithms deployed at the edge \cite{nikouei2018intelligent}. The tracking algorithm will record trajectory of each detected object by extracting useful information from each frame. Because of the resource constraint at edge devices, computing intensive tasks, such as feature extraction and suspicious behavior recognition, are outsourced to more powerful nodes like fog node or even to cloud centers.

\begin{figure}[t]
    \centering
        \includegraphics[width=0.48\textwidth]{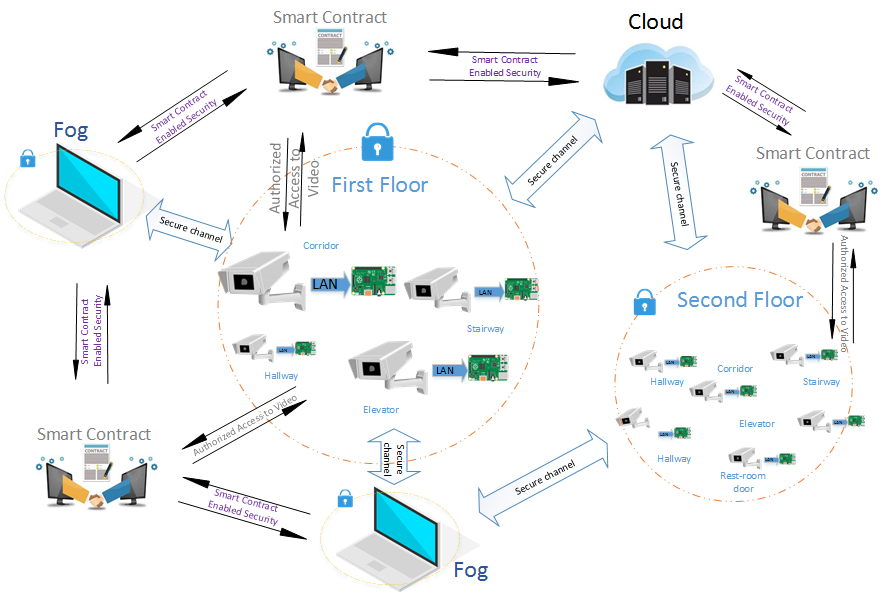}
    \caption{Illustration of the smart surveillance based on microservices architecture using smart Contract.}
    \label{fig:arch}
    \vspace{-10pt}
\end{figure}

The lower level tasks like object detection and tracking could be developed as microservices and deployed on edge devices, which are geographically close to each other and managed by a local fog node. Figure \ref{fig:arch} shows that all extracted information at the edge devices of each domain will be aggregated on domain specific fog node, like front door or second door. The communication channels between edge nodes and fog nodes are protected using encryption algorithms like AES-RSA key exchange. The features collected by fog nodes will be used by decision making algorithms and stored for future references. The high level tasks, such as behavior reorganization and malicious intention detection, are performed at more powerful cloud nodes. A security strategy is needed to protect data from tampering with and control granting authorized access to data. Blockchain is adopted to store the data securely and access control mechanism is enforced through smart contracts.

Figure \ref{fig:arch} demonstrates a scenario in which fog and cloud computing have different access levels given the functionalities of each node in system. Consider a campus environment, each security camera and the corresponding edge device will cover a small portion of a building or outdoor area. A fog node manages couple of the edge devices at the same time, so that part of a building such as one floor is managed by the same fog device. The entire building data is accessible for a cloud node and at the same time, the corresponding server at the police station needs access to data and real-time footage. The access control policy could be transcoded into a smart contract and deployed on the blockchain network by the administrator. All the nodes could interact through the ABI interfaces provided by the smart contract to request access. As shown in Fig. \ref{fig:arch}, the cloud nodes need a superior access to all nodes in their corresponding network. In such an occasion, access control policies are transcoded to smart contracts, which are considered as microservers, to authorize whether or not certain node can access the features that are stored on the fog or read the raw, real-time footage from the edge device. 

\section{Conclusion}
\label{sec:conclusion}

In this position paper, we propose to integrate the microservices architecture with the blockchain technology to enhance smart surveillance systems. Typical surveillance video analysis functions, like face detection, audio analysis, behavioral analysis and license plate recognition, are often conducted in parallel and independently. Therefore, development of these algorithms are not platform or programming language dependent under the paradigm of microservices architecture. The capability of continuous development and continuous delivery allow a more flexible and adaptive smart surveillance system. Algorithms can be upgraded without interrupting the current services. To secure the data exchanged among microservices, the blockchain technology is adopted to enable the operator track the data and avoid data tampering. In addition, smart contracts record and assign the access permission to certain surveillance data provided by a microservice dedicated to a sector. Smart contracts have automated the functioning of blockchain data and it provides the highest level of data encryption for efficient and secure communication. The sectors managed by the microservices includes extracting the video features and store it in a sector-specific database.


\ifCLASSOPTIONcaptionsoff
  \newpage
\fi

\bibliographystyle{IEEEtranS}

\bibliography{L-CNN.bib}

%





\end{document}